\documentclass[12pt]{article}
\usepackage{amssymb,amsmath,amsthm,amscd,latexsym}
\usepackage{mathrsfs}
\usepackage{mathrsfs}
\usepackage{amsfonts}
\usepackage{amsmath}
\usepackage{amssymb}
\usepackage{amscd}
\usepackage[all]{xy}
\usepackage{graphicx}
\usepackage{epstopdf}
\usepackage{color}

\renewcommand{\paragraph}{\roman{paragraph}}
\input cyracc.def

\newcommand{\Z}{\mathbb{Z}}
\newcommand{\F}{\mathbb{F}}

\newtheorem{thm}{\scshape \mdseries  Theorem}[section]
\newtheorem{lem}[thm]{\scshape \mdseries  Lemma}

\begin{document}
\title{\bf  Two-weight codes \\over\\ the integers modulo a  prime power
\thanks{This research is supported by National Natural Science Foundation of China (61672036) and by the Research Council  of Norway (247742/O70)}
}
\author{\small{Minjia Shi}\\ \small{School of Mathematical Sciences of Anhui University, China},
\and \small{Tor Helleseth}\\ \small{The Selmer center, Department of Informatics, University of Bergen, Bergen, Norway},
\and \small{Patrick Sol\'e}\\ \small{I2M, CNRS, Centrale Marseille, University of Aix-Marseille, Marseille, France}
}
\date{}
\maketitle
\begin{abstract}
 Let $p$ be a prime number. Irreducible cyclic codes of length $p^2-1$ and dimension $2$ over the integers modulo $p^h$ are shown to have exactly two nonzero Hamming weights. The construction uses the Galois ring of characteristic $p^h$ and order $p^{2h}.$ When the check polynomial is primitive,
 the code meets the Griesmer bound of (Shiromoto, Storme) (2012). By puncturing some projective codes are constructed. Those in length $p+1$ meet a Singleton-like bound of (Shiromoto , 2000). An infinite family of strongly regular graphs is constructed as coset graphs of the duals of these projective codes. A common cover of all these graphs, for fixed $p$, is provided by considering the Hensel lifting of these cyclic codes over the $p$-adic numbers.
\end{abstract}

{\noindent}{\bf Keywords:}   $2$-weight codes; Irreducible cyclic codes, strongly regular graphs
\section{Introduction}
Since the seminal paper of Delsarte \cite{D} two-weight codes have been studied in conjunction with combinatorial objects like strongly regular graphs (SRGs) \cite{BH,BM,D} or geometrical structures like caps in projective spaces \cite{CK}.
An important class of codes for construction of two-weight codes is that of irreducible cyclic codes \cite{BM,MR}. In particular it is conjectured that all two-weight projective irreducible cyclic codes are known \cite{SW,V2}.
In the special case of cyclic codes of dimension $2,$ it was shown that all of them (irreducible or not) are either one-weight or two-weight codes \cite{SZS,V}.
The connection between SRG's and two-weight codes over finite fields was extended recently to two-weight codes over rings for the homogeneous weight \cite{Byrne1, Byrne2}.

In the present paper, we take a third path, and construct two-weight codes over rings, but for the Hamming distance. In particular, we consider the ring of integers modulo $p^h,$ for some integer $h>1.$ Our construction is based on the Hensel lift over $\Z_{p^h}$ of irreducible cyclic codes
of dimension $2$ over the finite field $\F_p.$ The basic algebraic tool is the Galois ring of characteristic $p^h$ and size $p^{2h}.$
It turns out that different weight distributions are obtained, depending on the primitive status of the check polynomial. When the check polynomial is primitive, the codes meet  a Griesmer-like bound over rings \cite{SS}. To make the paper more readable, a complete proof of the weight distribution is only given for $h=3,$ and omitted for $h>3.$ The immediate but tedious generalization is left to the reader.
By suitable puncturing, we construct projective two-weight codes in length $p+1$ in the primitive case, and in lower length otherwise. The codes of length $p+1$ are optimal
 as Maximum Distance Rank codes \cite{SAS}. The main parameters of the SRG's  attached  to these two classes of codes via the coset graph construction on he dual code are given. They are of Latin square type \cite[p.121]{BH} in the primitive case. The strongly regular graphs so constructed, when $h$ varies, have a common cover which is a coset graph of a code defined on the $p$-adic integers, as those described in \cite{CS,ST}.

The paper is organized as follows. The next section collects the necessary notation and definitions. Section 3 is preliminary. Sections 4 and 5 are written for $h=3.$ Section 4 focusses on the weight distribution when the check polynomial is primitive,
while Section 5 considers the more general situation of an irreducible check polynomial of degre $2.$ Section $6$ construct projective codes from the codes with a primitive check polynomial, and describes the attached SRGs. Section $7$ generalizes the preceding three sections to an arbitrary $h.$
Section $8$ is the Conclusion.
\section{Definitions and Notation}

\subsection{Rings}
Throughout the paper, let $p$ be an odd prime. $\mathbb{Z}_{p^h}$ is the ring of integers modulo $p^h.$
The Galois ring $\mathcal{R}=GR(p^h,r)$ of order $p^{rh}$ and characteristic $p^h$ is the Galois extension of $\mathbb{Z}_{p^h}$ with degree $r$. It is a local ring, with maximal ideal $(p).$
The {\it Teichm\"uller} set $ \mathcal{T}=\{x \in GR(p^h,r)|x^{p^{r}}=x\}$ is a set of representatives of the {\it residue field $\mathbb{F}_{p^r}\simeq GR(p^h,r)/(p).$}
 If $z\in GR(p^s,m)$, let $\tilde{z}$ denote its image in $\mathbb{F}_{p^h}$ by reduction modulo $(p).$ This map is extended in the obvious way to the polynomial ring $GR(p^s,m)[x].$
It is known that $GR(p^h,r)=\mathcal{T}\oplus p\mathcal{T}\oplus \cdots\oplus p^{h-1}\mathcal{T} $
( $p$-adic expansion of $GR(p^h,r)$).
If $t \in \mathcal{T},$ define its conjugate as $t^p.$ For a general element $z$ of $GR(p^h,r)$ denote by $F(z)$ (or $\overline{z}$ if $r=2$) its conjugate obtained by linearity from its $p$-adic expansion.
The {\it trace} $Tr(z)$ from $GR(p^h,r)$ down to $\mathbb{Z}_{p^h}$ is then defined as the sum $$Tr(z)=\sum_{i=0}^{r-1}F^i(z).$$
\subsection{Codes over rings}
The {\em Hamming weight} of ${\bf x}\in \mathbb{Z}_{p^h}^n$ is denoted by $w_H({\bf x}).$
A linear {\it code} of length $n$ over $\mathbb{Z}_{p^h}$ is a submodule of $\mathbb{Z}_{p^h}^n$.
The {\em weight distribution} of a code $C$ of length $n$ over $\mathbb{Z}_{p^h}$ is defined as the list of lists
$$[<0,1>, \dots,<w_i,A_i>,\dots, <w_n,A_n> ],  $$
where $A_i$ is the numbers of $c \in C$ with $W_H(c)=i.$ The number $A_i$ is called the {\em frequency} of the weight $i.$
The {\it dual} $C^\bot$ of $C$ is understood with respect to the standard inner product. The {\em minimum distance} of a linear code is its minimum nonzero Hamming weight.
A linear code is {\em projective} if its dual has minimum distance $\ge 3.$

  A code is {\it cyclic} if it is linear and invariant under the shift.
 We consider cyclic codes of the form $\langle g(x)\rangle$ with $g(x)$ a divisor of $x^n-1.$ All the cyclic codes in this paper are {\it irreducible }, in the sense
 that their {\it check polynomial} $H(x)=\frac{x^n-1}{g(x)}$ is {\it basic irreducible} in $\mathbb{Z}_{p^h}[x]$ (that means that $H$ is monic and $\tilde{H}$ is irreducible in $\F_p[x]$) \cite{WZX}. Similarly, $H(x)$ will be said to be {\it primitive} if $\tilde{H}$ is primitive in $\F_p[x].$
 As is well-known if $H(x)$ is basic irreducible, and $H(\alpha)=0,$ then the code is a trace code of the form
 $$C=\{c(A)=\big(Tr(A\alpha^i)\big)_{i=0}^{n-1} \mid A \in \mathcal{R}\}.$$
 Note that if $\alpha$ has order $b$ then $|\{\alpha^i \mid i=0,\dots,n-1\}|=b.$
 The parameters of a two-weight code $C$ over an alphabet $A$ of size $q$ are listed as $[n,k,\{w_1,w_2\}]_q$ if $A$ is a finite field, and $C$ is of dimension $k,$  and $(n, |C|, \{w_1,w_2\})_q$ if $A$ is not a finite field.
 \subsection{Graphs}
 A simple graph on $v$ vertices is called a {\em
    strongly regular graph} with parameters $(v, \eta, \lambda, \mu)$ if
  \begin{enumerate}
  \item each vertex is adjacent to $\eta$ vertices;
  \item for each pair of adjacent vertices, there are $\lambda$
    vertices adjacent to both;
    \item for each pair of non-adjacent vertices, there are $\mu$
      vertices adjacent to both.
  \end{enumerate}

An {\em eigenvalue} of a graph $\Gamma$ (i.e., an eigenvalue of its adjacency
matrix) is called a \emph{restricted eigenvalue} if there is a
corresponding eigenvector which is not a multiple of the all-one vector
\textbf{1}. Note that for an $\eta$-regular connected graph, the
restricted eigenvalues are simply the eigenvalues different from $\eta$.

The {\em coset graph} of a projective code $C\subseteq \mathbb{Z}_{p^h}^n$ has for vertices the cosets of $C$ , two vertices being connected iff they differ by a coset of minimum Hamming weight one.
\section{Preliminaries}
Let $R=\Z_p^h.$ Denote by $\mathcal{R}=GR(p^h,2),$ a quadratic extension of $R,$ and by $Z= \mathcal{T}\setminus \{0\}$ . Thus $|Z|=p^2-1.$
We consider trace codes defined by
$$C_d=\{c(A)=\big(Tr(Ax)\big)_{x\in Z^d} \mid A \in \mathcal{R}\},$$
where $d$ is an arbitrary divisor of $p^2-1,$ and $Z^d$ is the multiset $\{\{ x^d \mid x \in Z\}\},$ repetitions being allowed. In fact $Z^d$ is the repetition of $d$ sets of size $\frac{p^2-1}{d}.$
By the preceding section, we see that  $C_1$ is permutation equivalent to a cyclic code with a primitive check polynomial, while $C_d$ in general, is permutation equivalent to a cyclic code with a basic irreducible check polynomial the roots of which have order $\frac{p^2-1}{d}.$
We will use the following Lemma.

{\lem \label{wt} $$w_H(c(A))=p^2-1-|\{x \in Z \mid Ax^d+\overline{A}x^{dp}=0\}|.$$}

\begin{proof}
Because $\mathcal{R}$ is a quadratic extension of $R,$ the Trace is a sum of two terms. $Tr(Ax^d)=Ax^d+\overline{A}x^{dp}.$
The weight of $c(A)$ is the length minus the number of times $Tr(Ax)=0,$ for $x \in Z^d.$
\end{proof}
\section{Primitive  check polynomial}
In this section, we determine the weight distribution of $C_1.$

{\thm \label{prim} The code $C_1$ is a two-weight code with $w_1=p^2-p,$ and $w_2=p^2-1.$ Letting $A_1,\, A_2$ denote their respective frequencies we have
\begin{eqnarray}
A_1&=&(p+1)(p^3-1),\\
A_2&=&p(p^2-1)(p^3-1).
\end{eqnarray}
}
\begin{proof} In view of Lemma \ref{wt}, we need to count the solutions in $x \in Z$ of $Tr(Ax)=0.$
Write $A=a+pb+p^2c,$ with $a,b,c \in Z\cup \{0\}.$ The equation $Tr(Ax)=0$ can be rewritten as

\begin{equation}\label{padic} ax+pbx+p^2cx=-\big(a^px^p+pb^px^p+p^2c^px^p\big) .\end{equation}

We remark that, if $p$ is odd, we have $(-1)\in Z,$ since $(-1)^{p^2-1}=1.$ Thus the terms $-a^px^p,\,-b^px^p,\,-c^px^p$ are in $Z,$ just like $ax,\,bx,\,cx.$

By unicity of the $p$-adic expansion in $GR(p^3,2),$  equation (\ref{padic}) yields the system

\begin{eqnarray}\label{sys}
ax&=&-a^px^p,\\
bx&=&-b^px^p,\\
cx&=&-c^px^p.
\end{eqnarray}

Up to permutations of $a,b,c,$ we claim that three cases can occur where this system has at least one solution. In each case the number of solutions turns out to be $p-1.$

\begin{enumerate}
\item \underline{ $a\neq 0, b=c=0$} There are $p-1$ solutions of $x^{p-1}=-a^{1-p},$ if $-1=\epsilon ^{p-1},$ for some $\epsilon \in Z.$ This is possible if $(-1)^{p+1}=1,$ which holds true for $p$ odd.
\item \underline{ $ab\neq 0, c=0$} There are $p-1$ solutions of $x^{p-1}=-a^{1-p}=-b^{1-p},$ provided $a^{p-1}=b^{p-1}.$

\item \underline{ $abc\neq 0$} There are $p-1$ solutions of $x^{p-1}=-a^{1-p}=-b^{1-p}=-c^{1-p},$ provided $a^{p-1}=b^{p-1}=c^{p-1}.$
\end{enumerate}

The number of values of $A$ for each case is

\begin{enumerate}
\item \underline{ $a\neq 0, b=c=0$} $p^2-1$ since $a$ is arbitrary in $Z.$
\item \underline{ $ab\neq 0, c=0$} $(p^2-1)(p-1)$ since $(\frac{b}{a})^{p-1}=1.$
\item \underline{ $abc\neq 0$} $(p^2-1)(p-1)^2$ since $(\frac{b}{a})^{p-1}=(\frac{c}{a})^{p-1}=1.$

Thus, accounting for permutations of $a,b,c$ we obtain $$A_1=3(p^2-1)+3(p^2-1)(p-1)+(p^2-1)(p-1)^2=\frac{(p^2-1)}{p-1}(p^3-1)=(p+1)(p^3-1).$$

Since $1+A_1+A_2=|C|=p^6,$ the result follows.
\end{enumerate}
\end{proof}

{\bf Examples:}
\begin{itemize}
\item When $p=5,$ we obtain codes of length $24,$ with the following weight distributions.\\
$[ <0, 1>, <20, 744>, <24, 14880> ] $\\
\item When $p=7,$ we obtain codes of length $48,$ with the following distinct weight distributions.\\
$ [ <0, 1>, <42, 2736>, <48, 114912> ]$\\
\item When $p=11,$ we obtain codes of length $48,$ with the following distinct weight distributions.\\
$ [ <0, 1>, <110, 15960>, <120, 1755600> ].$
\end{itemize}

\section{Irreducible check polynomial}
In this section $d>1.$ Let $(x,y)$ denote the GCD of the integers $x$ and $y.$
{\thm\label{primd} The code $C_d$ is a two-weight code with $w_1=p^2-1-m,$ where $m=(d,p+1)(p-1),$ and $w_2=p^2-1.$ Letting $A_1,\, A_2$ denote their respective frequencies we have
\begin{eqnarray}
A_1&=&\frac{(p^2-1)}{m}((m+1)^3-1),\\
A_2&=&p^6-1-A_1.
\end{eqnarray}
}

\begin{proof} (sketch)
The proof is similar to that of Theorem \ref{prim}. The system of equations (4),(5),(6) is replaced by
\begin{eqnarray}\label{sys}
ax^d&=&-a^px^{dp},\\
bx^d&=&-b^px^{dp},\\
cx^d&=&-c^px^{dp}.
\end{eqnarray}

Because $x\in Z$, we have $x^{d(p-1)}=x^m,$ where $$m=(d(p-1),p^2-1)=(d,p+1)(p-1).$$ The same discussion as in the proof of Theorem \ref{prim} yields
$$A_1=3(p^2-1)+3(p^2-1)m+(p^2-1)m^2=\frac{(p^2-1)}{m}((m+1)^3-1).$$
\end{proof}

{\bf Examples:} If $m=p-1$ the weight distribution is the same as in the primitive case.
In the case $m>p-1,$ the following values were computed in Magma \cite{M}.\\
\begin{itemize}
\item When $p=5,$ we obtain codes of length $24,$ with two distinct weight distributions when $d>1$ varies.\\
$[ <0, 1>, <12, 248>, <24, 15376> ],\,$
$[ <0, 1>, <16, 372>, <24, 15252> ].$\\
\item When $p=7,$ we obtain codes of length $48,$ with two distinct weight distributions when $d>1$ varies.\\
$[ <0, 1>, <24, 684>, <48, 116964> ],\,$\\
$[ <0, 1>, <36, 1368>, <48, 116280> ].$\\
\item When $p=11,$ we obtain codes of length $48,$ with four distinct weight distributions when $d>1$ varies.\\
$[ <0, 1>, <60, 2660>, <120, 1768900> ],\,$\\
$[ <0, 1>, <80, 3990>, <120, 1767570> ],$\\
$[ <0, 1>, <90, 5320>, <120, 1766240> ],\,$\\
$[ <0, 1>, <100, 7980>, <120, 1763580> ].$
\end{itemize}

\section{Projective codes and SRG's}
\subsection{Projective codes}
If $C$ is a linear code over $\Z_{p^3},$ we denote by $\widehat{C}$ the projective code obtained by removing linearly dependent coordinates.

{\thm\label{param} The  code $\widehat{C_1}$ is a two-weight code with parameters\\ $(p+1,p^6,\{p,p+1\})_{p^3}.$
It is optimal with these parameters.
}

\begin{proof}
Note that two columns $x,y \in Z$ of the generator matrix of $C_1^\bot$ are linearly dependent iff $x/y \in Z\cap \Z_{p^3},$ iff $(x/y)^{p-1}=1.$
Thus, the parameters of $\widehat{C_1}$ are obtained from those of ${C_1}$ by dividing the length and the weights by $p-1.$
This code meets the Singleton bound of \cite[Chap, 12]{SAS} and \cite{Sh}. Indeed, it is a free code of rank $2$, length $p+1$ and distance $p.$
It is thus MDR in the sense of \cite[Chap, 12]{SAS}.
\end{proof}

The analogous Theorem for $d>1$ is as follows.
{\thm\label{paramd} The  code $\widehat{C_d}$ is a two-weight code with parameters\\ $(\frac{p^2-1}{m},p^6,\{\frac{p^2-1}{m}-1,\frac{p^2-1}{m}\})_{p^3},$ where $m=(d,p+1)(p-1).$
}
\begin{proof}
Note that two columns labelled by $x,y \in Z$ of the generator matrix of $C_d^\bot$ are linearly dependent iff $x^d/y^d \in Z\cap \Z_{p^3},$ iff $(x/y)^{d(p-1)}=1.$
Now $(x/y)^{d(p-1)}=(x/y)^{m}$ where $m=(d(p-1),p^2-1)=(d,p+1)(p-1).$
Thus, the parameters of $\widehat{C_d}$ are obtained from those of ${C_d}$ by dividing the length and the weights by $m.$
\end{proof}

\subsection{Their graphs}
{\thm\label{SRG} The coset graph of $\widehat{C_1}^\bot$ is a SRG of degree $(p+1)(p^3-1),$ on $p^6$ vertices with restricted eigenvalues $p^3-p-1$ and $-(p+1)$ of respective multiplicities $A_1$ and $A_2$ of Theorem \ref{prim}.}
\begin{proof} By Theorem 11.1.11 of \cite{BCN}
the restricted eigenvalues are computed as $\lambda_i=n(p^3-1)-pw'_i$ for $i=1,2$ with the weights $w'_1=p$ and $w'_2=p+1$ from Theorem \ref{param}, and their multiplicities equal the frequency of the corresponding weights.
\end{proof}

{\bf Example:} We take $p=3$ to obtain an SRG on $729$ vertices of degree $104,$ and eigenvalues $23$ and $-4$ with respective multiplicities
$104$ and $624.$ As per \cite{BT}, alternate constructions include a $[52,6,\{27,36\}]_3,$ and a $[13,3,\{9,12\}]_9.$

{\thm\label{SRGd} The coset graph of $\widehat{C_d}^\bot$ is a SRG of degree $\frac{(p^2-1)}{m}(p^3-1),$ on $p^6$ vertices with restricted eigenvalues $p^3-\frac{(p^2-1)}{m}$ and $-\frac{(p^2-1)}{m}$ of respective multiplicities $A_1$ and $A_2$ of Theorem \ref{primd}.}

\begin{proof} By Theorem 11.1.11 of \cite{BCN}
the restricted eigenvalues are computed as $\lambda_i=n(p^3-1)-pw'_i$ for $i=1,2$ with the weights $w'_1=\frac{p^2-1}{m}-1$ and $w'_2=\frac{p^2-1}{m}$ from Theorem \ref{paramd}, and their multiplicities equal the frequency of the corresponding weights.
\end{proof}

\section{Generalization}
\subsection{Codes}
We give without proof the generalization of Theorem \ref{prim} from $GR(p^3,2),$ to $GR(p^h,2).$

{\thm \label{prim} The code $C_1$ is two-weight code with $w_1=p^2-p,$ and $w_2=p^2-1.$ Letting $A_1,\, A_2$ denote their respective frequencies we have
\begin{eqnarray}
A_1&=&(p+1)(p^h-1),\\
A_2&=&p(p^{h-1}-1)(p^h-1).
\end{eqnarray}
}

This code is optimal as the next result shows.

{\thm The code $C_1$ meets the Griesmer bound for QF rings with equality.}

\begin{proof}
Note that the residue field of $\Z_{p^h}$ is $\F_p.$ The rank of the free code $C_1$ is $2$ and its minimum Hamming distance $p^2-p.$
By the theorem 3.11 of \cite[p.26]{SAS}, or \cite{SS}, we know that its length
$$n\ge (p^2-p)+\lceil \frac{p^2-p}{p} \rceil=p^2-1.$$
But, by construction $n=p^2-1.$ The result follows.
\end{proof}

{\thm\label{wh} Assume $d>1,$ and $h>1.$ The code $C_d$ is a two-weight code with $w_1=p^2-1-m,$ where $m=(d,p+1)(p-1),$ and $w_2=p^2-1.$ Letting $A_1,\, A_2$ denote their respective frequencies we have
\begin{eqnarray}
A_1&=&\frac{(p^2-1)}{m}((m+1)^h-1),\\
A_2&=&p^{2h}-1-A_1.
\end{eqnarray}
}

{\bf Remark:} If $h=1,$ and $m=p-1,$ we have $A_1=p^{2}-1,$ and $C_d$ is a one-weight code. This is the case $u=1$ of \cite{V}.

{\thm The  code $\widehat{C_1}$ is a two-weight code with parameters $$(p+1,p^{2h},\{p,p+1\})_{p^h}.$$ It is optimal with these parameters.
}
\begin{proof}
The parameters of $\widehat{C_1}$ are obtained from those of ${C_1}$ by dividing length and weights by $p-1.$
This code meets the Singleton bound of \cite[Chap, 12]{SAS}. Indeed, it is a free code of rank $2$, length $p+1$ and distance $p.$
It is thus MDR in the sense of \cite[Chap, 12]{SAS}.
\end{proof}
The analogous result for $d>1$ is as follows.
{\thm The  code $\widehat{C_d}$ is a two-weight code with parameters $$(\frac{(p^2-1)}{m},p^{2h},\{\frac{(p^2-1)}{m}-1,\frac{(p^2-1)}{m}\})_{p^h}.$$
}
{\bf Example:} With $p=7,\, h=2,\, d=2,\, m=12$ the code $\widehat{C_2}$ has length $4$ and weight distribution $[ <0, 1>, <2, 96>, <4, 2304> ].$
\subsection{Finite Graphs}
The proof of the following Theorem is analogous to that of Theorem \ref{SRG} and is omitted.
{\thm The coset graph of $\widehat{C_1}^\bot$ is a SRG of degree $(p+1)(p^h-1),$ on $p^{2h}$ vertices with restricted eigenvalues $p^h-p-1$ and $-(p+1)$ with respective multiplicities $A_1$ and $A_2$ of Theorem \ref{wh}.}

{\bf Example:} With $p=2,\, h=4,$ we obtain a SRG on $256$ vertices, degree $45,$ unrestricted eigenvalues $13$ and $-3.$
Alternate constructions include as per \cite{BT} a $[15,4,\{8,12\}]_4,$ a binary $[45,8,\{16,24\}].$

{\bf Remark:} In \cite[p.121]{BH} an SRG is said to be of Latin square type if its parameters are
$$(v, \eta, \lambda, \mu) = (N^2, M(N - 1),(M - 1)(N - 2) + N - 2, M(M - 1)),$$ for some integers $M,N$  with restricted eigenvalues
$N-M,\,-M,$ and respective multiplicities $M(N-1)$ and $(N-M+1)(N-1).$  It can checked that the parameters above are of this form with $N=p^h,\,M=p+1.$

The proof of the following is analogous to that of Theorem \ref{SRGd} and is omitted.
{\thm The coset graph of $\widehat{C_d}^\bot$ is a SRG of degree $\frac{(p^2-1)}{m}(p^h-1),$ on $p^{2h}$ vertices with restricted eigenvalues $p^h-\frac{(p^2-1)}{m}$ and $-\frac{(p^2-1)}{m}$ with respective multiplicities $A_1$ and $A_2$ of Theorem \ref{wh}.}

{\bf Example:} With $p=7,\, h=2,\, d=2,$ the code $\widehat{C_2}$ has length $4$ and we obtain a SRG on $7^4=2401$ vertices of degree $192$ with restricted eigenvalues $94,\, -4.$
These parameters are beyond the table of \cite{BT}. They are not of Latin square type (see the preceding Remark).
\subsection{Infinite Graphs}
Denote by $\Gamma_h$ the coset graph of $\widehat{C_1}^\bot$ over $\Z_{p^h}.$
Following \cite{CS}, we denote by $\Z_{p^\infty},$ the ring of $p$-adic integers, that is to say the topological closure of $\Z$ for the $p$-adic topology \cite{S}.
Denote by $\Gamma_\infty$ the coset graph of $\widehat{C_1}^\bot$ over $\Z_{p^\infty}.$ Both $\widehat{C_1}$ and $\widehat{C_1}^\bot$ can be seen as obtained by extension of scalars from
their counterparts over $\F_p,$ or as Hensel lifts from them \cite{CS}. Thus $\Gamma_\infty$ is a graph with a denumerably many vertices.
Recall that a {\em cover} of a graph $H$ by a graph $G$ is an adjacency preserving surjection from $G$ to $H.$
The next result shows that, roughly speaking, $\Gamma_\infty$ is a kind of limit of the $\Gamma_h$'s.
{\thm For all $h>0,$ we have
\begin{itemize}
\item $\Gamma_{h+1}$ is a cover of $\Gamma_h,$
\item $\Gamma_\infty$ is a cover of $\Gamma_h.$
\end{itemize}}

\begin{proof}
Follows immediately by reduction modulo $p^h,$ that preserves the coset graph definition.
\end{proof}

A similar result holds for any fixed $d>1$ that divides $p^2-1.$
\section{Conclusion}
In this paper we have constructed two-weight codes over the rings $\Z_{p^h},$ by considering irreducible cyclic codes of dimension $2.$ This opens the way to considering other families of cyclic codes over these rings, or cyclic codes over other families of rings. The natural candidates for a different alphabet are chain rings, but less general choices might be more fruitful. Irreducible cyclic codes of dimension three or more over rings might have many weights.

We have used these special codes to construct SRGs. It seems, based on computations in this paper and in \cite{SZS}, that the coset graphs of the dual of an MDS code of dimension $2$ always give an SRG of Latin square type. There might be a direct combinatorial explanation to that fact, in view of the well-known equivalence between Mutually Orthogonal Latin Squares and MDS codes \cite{G}.

It is a worthwhile project  to compute the spectrum of the graph $\Gamma_{\infty}$ defined as the spectrum of its adjacency operator. An engineering application can be found in \cite{ST}.

\end{document}